\documentclass[preprint2]{aastex62}
\usepackage{natbib}
\usepackage{hyperref}
\usepackage{amsmath}
\usepackage[mathlines]{lineno}
\usepackage{multirow}
\usepackage{bm}
\usepackage{graphicx}
\usepackage{longtable}
\usepackage{xcolor}

\begin{document}

\title{Mass function of stellar black holes as revealed by the LIGO-Virgo-KAGRA observations }

\author{Xiao-Fei Dong}
\affiliation{School of Astronomy and Space Science, Nanjing University, Nanjing 210023, China}
\author{Yong-Feng Huang}\thanks{Email:hyf@nju.edu.cn}
\affiliation{School of Astronomy and Space Science, Nanjing University, Nanjing 210023, China}
\affiliation{Key Laboratory of Modern Astronomy and Astrophysics (Nanjing University), Ministry of Education, Nanjing 210023, China}
\author{Zhi-Bin Zhang}
\affiliation{School of Physics and Engineering, Qufu Normal University, Qufu 273165, China}
\author{Xiu-Juan Li}
\affiliation{School of Cyber Science and Engineering, Qufu Normal University, Qufu 273165, China}
\author{Ze-Cheng Zou}
\affiliation{School of Astronomy and Space Science, Nanjing University, Nanjing 210023, China}
\author{Chen-Ran Hu}
\affiliation{School of Astronomy and Space Science, Nanjing University, Nanjing 210023, China}
\author{Chen Deng}
\affiliation{School of Astronomy and Space Science, Nanjing University, Nanjing 210023, China}
\author{Yang Liu}
\affiliation{School of Physics and Engineering, Qufu Normal University, Qufu 273165, China}

\begin{abstract}

Ninety gravitational wave events have been detected by the
LIGO-Virgo-KAGRA network and are released in the
Gravitational-Wave Transient Catalog. Among these events, 83 cases
are definitely binary black hole mergers since the masses of all
the objects involved significantly exceed the upper limit of
neutron stars. The black holes in these merger events naturally
form two interesting samples, a pre-merger sample that includes
all the black holes before the mergers and a post-merger sample
that consists of the black holes generated during the merging
processes. The former represents black holes that once existed in
the Universe, while the latter represents newly born black holes.
Here we present a statistical analysis on these two samples. The
non-parametric $\tau$ statistic method is adopted to correct for
the observational selection effect. The Lynden-Bell's $C^{-}$ method
is further applied to derive the mass distribution and density
function of black holes. It is found that the mass distribution
can be expressed as a broken power-law function. More
interestingly, the power-law index in the high mass region is
comparable for the two samples. The number density of black holes
is found to depend on redshift as $\rho(z) \propto z^{-2.06}$ --
$z^{-2.12}$ based on the two samples. Implications of these findings
on the origin of black holes are discussed.

\end{abstract}

\section{Introduction}

As the first gravitational wave (GW) event being detected, GW150914 was produced by
the merger of two black holes (BHs) whose masses are $m_{1} =36^{+5}_{-4}M_{\odot}$
and $m_{2} = 29^{+4}_{-4}M_{\odot}$, respectively \citep{2016PhRvL.116f1102A}.
Since then, the LIGO-Virgo-KAGRA (LVK) gravitational wave detector
network \citep{2015CQGra..32g4001L, 2015CQGra..32b4001A, 2018PTEP.2018a3F01A} has
recorded about 90 confident binary merger events, which are reported in the Gravitational-Wave
Transient Catalog (GWTC) \footnote{https://gwosc.org/} \citep{2021PhRvX..11b1053A, 2023PhRvX..13d1039A, 2023PhRvX..13a1048A}.
While the possibility that neutron stars might be involved in four or five GW events still cannot
be completely expelled, it is believed that the majority of the GW events were produced by
binary black hole (BBH) mergers. In fact, among the 90 GW events, at least 83 candidates
come from BBH systems with the black hole mass ranging
in  5 -- 140 $M_{\odot}$ \citep{2023MNRAS.526.4908C, 2023PhRvX..13a1048A}.
These GW events thus provide a valuable sample of stellar mass black holes.

GW observations provide useful information on the masses and
distances of the black hole members, which can be used to probe
the binary origin. There are many opinions on the origin of binary
black holes, such as binaries in the field area of galaxies with a
relatively low stellar density \citep{1998ApJ...506..780B,
2018MNRAS.480.2011G, 2021ApJ...922..110G, 2024ApJ...973..132A},
dynamically-driven BBHs in dense stellar clusters
\citep{1993Natur.364..421K, 2000ApJ...528L..17P,
2010MNRAS.402..371B, 2016PhRvD..93h4029R, 2022MNRAS.513.4527C},
BBHs originated from triple systems \citep{2017ApJ...841...77A,
2020ApJ...903...67M, 2021ApJ...907L..19V}, gas capture in the
disks of active galactic nuclei \citep{2012MNRAS.425..460M,
2017ApJ...835..165B, 2019MNRAS.488.2825F, 2020ApJ...899...26T}.
BBHs might also originate as part of a primordial
BH population in the early universe \citep{1974MNRAS.168..399C,
2016PhRvL.116t1301B, 2017PhRvD..96l3523A,
2020PhRvD.101h3008W,
2022ApJ...931L..12N, 2023JCAP...03..024C}.
The exact processes that give birth to the BBHs detected by LVK
have not yet been conclusively determined
\citep{2024ApJ...961....8F}, but note that different channels
could lead to different BBH characteristics, such as their
distributions in mass, spin, distance and other parameters.

Many researchers have tried to infer the
physics of binary mergers and the origins of BBHs by studying the
distributions of BBH parameters. The parametric method, such as
the Bayesian analysis, has been widely applied to a population of
BBHs to get the distribution of their physical parameters
(population-level, \citealt{2014ApJ...795...64F,
2016PhRvX...6d1015A, 2017Natur.548..426F, 2018ApJ...863L..41F,
2021ApJ...914L..34P, 2024PhRvD.109h4056S}). The parameters of a
certain BBH can even be inferred in some cases by using this
method (event-level, \citealt{2017MNRAS.465.3254M,
2020ApJ...891L..31F, 2022PhRvD.105d4034G, 2024ApJ...962..169E,
2024arXiv240611885R}). Another popular method is the
non-parametric model which does not rely on any pre-assumptions of
the parameter distributions \citep{2019MNRAS.486.1086M,
2022PhRvD.105l3014S, 2023ApJ...946...16E, 2024PhRvX..14b1005C,
2024arXiv240616813H}. For example, \cite{2020ApJ...891L..31F} used
both parametric and non-parametric method to derive the
mass-weighted BBH event rate. It is found that, in contrast to the
cut-off feature in the result of the parametric model, the
non-parametric model tends to extrapolate smoothly to high masses
and provides a conservative upper limit on the rate of high-mass
mergers. A semi-parametric method that incorporates a simple
parametric component with an additional non-parametric component
is also applied to obtain the distribution of BBH parameters
\citep{2022ApJ...924..101E, 2023ApJ...955..107F, 2024PhRvL.133e1401L}. It explains most of the structure through
parametric models while retaining the flexibility of the
non-parametric part.

The number of BHs existed in a unit comoving volume, i.e., the comoving BH number density, can provide information on the number of BHs formed at a certain redshift and can help to understand the
progenitors of BHs at various stages of evolution
\citep{2002ApJ...574..554L}. However, there are several selection
effects in the GW observations of BBHs \citep{2019RNAAS...3...66F,
2024ApJ...962..169E}, which prevent us from
deriving the redshift distribution and mass distribution of BHs
directly. As a result, the number densities of BHs in BBH systems
could be derived only when the selection effects are properly
accounted for.

The Lynden-Bell's $C^{-}$ method \citep{1971MNRAS.155...95L} is usually used to solve mutually
independent truncated bivariate data distributions. It has been applied in many fields of
astronomy, such as short/long gamma-ray
bursts \citep{2002ApJ...574..554L, 2015ApJS..218...13Y, 2016A&A...587A..40P, 2018ApJ...852....1Z, 2021RAA....21..254L, 2022MNRAS.513.1078D, 2023ApJ...958...37D, 2024MNRAS.527.7111L},
fast radio bursts \citep{2019JHEAp..23....1D},
galaxies \citep{1978AJ.....83.1549K, 1986ApJ...307L...1L, 1986MNRAS.221..233P} and
quasars \citep{2011ApJ...743..104S, 2021ApJ...913..120Z}. The method is quite useful in
deriving the intrinsic luminosity functions of various objects based on their flux and redshift
measurements. Interestingly, it is also proved to be effective in correcting for the observational
selection effects whenever a bivariate (or, more generally, multivariate) distribution is
involved \citep{1971MNRAS.155...95L, 1992ApJ...399..345E, 1998astro.ph..8334E, 2015MNRAS.451.3898D, 2022ApJ...925...15L}.

A key assumption in the $C^{-}$ method is about the data
independence. Therefore, it is important to ensure the
independence of the truncated data. The $\tau$ statistic method is
a unique non-parametric technique widely applied to truncated data
for assessing the independence of parameters
\cite{1992ApJ...399..345E}. The joint operation of the $\tau$
statistic method and Lynden-Bell's $C^{-}$ method provides an ideal
non-parametric method. It is extremely effective for a truncated
sample since it does not depend on any pre-assumed models and can
give a point-by-point description of the cumulative distribution.

In this study, we adopt the non-parametric method to explore the mass function and redshift
distribution of BHs associated with BBH mergers. The BH number density will also be derived
based on the analysis. The structure of our paper is organized as follows. In
Section~\ref{sec: DATA}, data acquisition and the two BH samples used for the analysis are
described. Section~\ref{sec: METHOD} introduces the non-parametric method and the calculation
processes in detail. The numerical results are presented in Section~\ref{sec: RESULTS}. Finally, we end up with our conclusions and discussion in Section~\ref{sec: Conclusion}.

\section{BBH merger events}
\label{sec: DATA}

The LVK gravitational wave detector network has completed three rounds of observation
operations (O1/O2/O3) as of March 27, 2020 \citep{2018LRR....21....3A, 2021PhRvX..11b1053A, 2023PhRvX..13d1039A}. It began the fourth
observing run on May 24, 2023. The online GWTC is a cumulative set of gravitational wave transients
maintained by the LVK collaboration.
It contains confirmed GW events with a
credible probability of being of astrophysical origin ($p_{astro}
> 0.5$), which are unlikely due to instrumental noise.
Totally 90 confident GW events are included in the catalog as
of March 27, 2020. Note that in 7 events, the mass of at least one
binary member is not massive enough \citep{2018LRR....21....3A} to
be definitely identified as a black hole.  In other words, these 7
events might be generated by binary neutron star mergers or
neutron star-BH mergers. Since we are only interested in BBH
mergers here, we exclude these GW events in our analyses.

Recently, the LVK collaboration specially reported a new GW event, GW230529\_181500, detected
during a preliminary analysis of the O4 data \citep{2024ApJ...970L..34A}. It seems to be produced
by the coalescing of a less massive compact binary, with the masses of the two objects ranging
in 2.5 -- 4.5 $M_{\odot}$ and 1.2 -- 2.0 $M_{\odot}$. As a typical lower mass gap system, it is
widely believed that the compact stars involved are neutron stars, although the possibility that
they are BHs still cannot be excluded \citep{2024JCAP...08..030H}. Considering this uncertainty,
we do not include the event in our study.

To sum up, we have 83 GW events in which all the compact objects are confirmed BHs.
Our sample is composed of these BHs.
Note that $~83\%$ of the BBHs in our sample have a false alarm rate (FAR)
less than 1 per year, and $~100\%$ of then have a $p_{astro}$
larger than 0.5.

For each GW event, we denote the mass of the
heavier companion as $m_{1}$ and the mass of the lighter BH as $m_{2}$. The mass of the
final product of the merger, i.e. the newly born massive BH, is denoted as $m_{f}$. Note that during the GW observation of a merger event, the unknown spin of each black hole will affect the estimated BH masses. In the GWTC catalog, the BH masses are estimated based on
the assumption that the BH spins are in accordance with the angular speed of the innermost
stable circular orbit. For a GW event, the chirp mass and mass ratio $q$ can be derived
from the waveform information. The values of $m_{1}$ , $m_{2}$  and $m_{f}$ can then be
determined from the chirp mass and $q$ \citep{1995PhRvD..52..848P, 2013ApJ...766L..14H}.
To assess the event density of BBH mergers, we need the redshift ($z$) information, which is
also available in the GWTC catalog \citep{2018LRR....21....3A}.
The relevant parameters of
these BHs have been taken from the GWTC website and are listed
in Table~\ref{tab:1} \citep{2016PhRvL.116f1102A, 2021PhRvX..11b1053A, 2023PhRvX..13d1039A}.

\begin{figure*}
     \centering
     \includegraphics[width=0.9\textwidth]{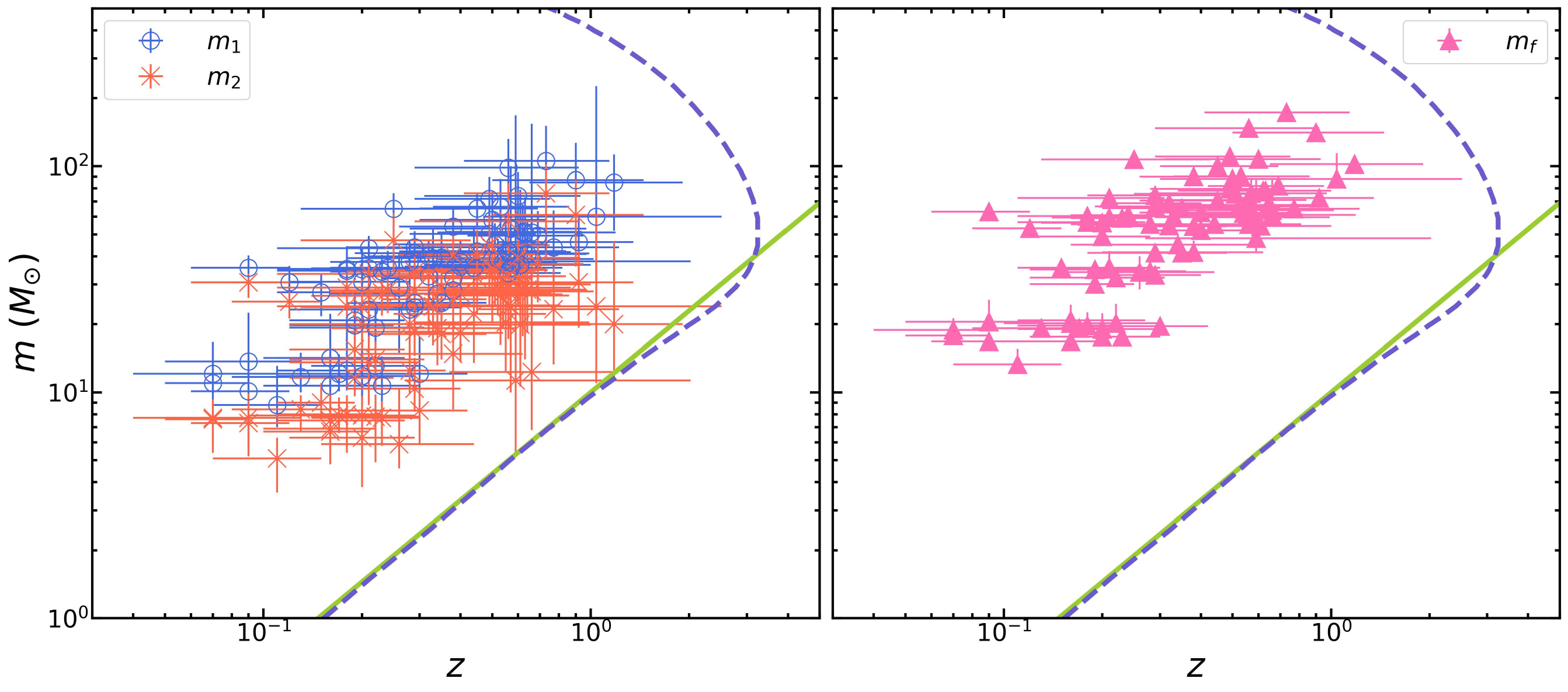}
     \caption{Distribution of the black hole masses and redshifts.
        Left panel: mass versus redshift for the pre-merger sample. The crosses and
        hollow circles represent the lighter ($m_2$) and heavier ($m_1$) black holes, respectively.
        Right panel: mass versus redshift for the post-merger sample.
        In both panels, the dashed line represents the truncation limit
        and the solid line is the baseline used in our statistical analysis. }
     \label{fig1}
\end{figure*}

All the confirmed BHs are further divided into two samples. The first sample is called the
pre-merger sample ($S_{pre}$), which includes all the separate black holes in the binary
systems before the merging process. So, the number of BHs in the $S_{pre}$ sample is
$83 \times 2 = 166$. It represents one part of the confirmed BHs once existed in the universe.
The second sample is called the post-merger sample ($S_{post}$), which includes all the
BHs produced after the merger. The number of BHs in the $S_{post}$ sample is simply 83.
It represents the newly born BHs observed in the universe, which can also be regarded
as a sample of stellar BHs at the relatively higher mass segment.
Figure~\ref{fig1} plots the distribution of the two samples of BHs on the mass-redshift plane.

In our study, we assume a flat $\Lambda$-CDM cosmology with the Hubble parameter
and the density parameters taken as $H_{0}=67.9~{\rm km~s^{-1}~Mpc^{-1}}$,
${\Omega}_{m}=0.3065$, and ${\Omega}_{\Lambda}=0.6935$. It is consistent with
the cosmology parameters taken by the LVK
collaboration \citep{2016A&A...594A..13P, 2018LRR....21....3A, 2023PhRvX..13d1039A}.

\section{The non-parametric method}
\label{sec: METHOD}

The mass function and number density of BHs can be inferred from the observed BH samples
given that the BH masses and redshifts are available. However, note that the BH samples are truncated
due to the detection threshold of the GW detectors.
Similar to previous studies, an event is regarded as being
credibly identified when the observed signal to noise ratio (SNR) is no less than 8.
In Figure~\ref{fig1}, the dashed line shows the truncation line,
which represents the maximum redshift of BHs with a certain mass
that the upgraded (A+) Advanced LIGO can detect when the
SNR is no less than 8
\citep{2021CQGra..38e5010C, 2021arXiv210909882E}. The
observational bias caused by this truncation line should be
considered when we try to reconstruct the intrinsic distributions
of BHs based on the observational data.

The BHs form a bivariate sample with two truncated parameters, i.e. mass ($m$) and
redshift ($z$). The joint distribution of BHs is a function of both mass and distance,
which can be written as $\Psi(m, z)$. If the dependence of the joint distribution on
the two parameters is independent of each other, then we have $\Psi(m, z) = \psi(m)\phi(z)$,
where $\psi(m)$ represents the mass function and $\phi(z)$ represents the redshift distribution
of black holes. The Lynden-Bell's $C^{-}$ method \citep{1971MNRAS.155...95L} is suitable for
dealing with such mutually independent bivariate problems, which can help obtain the number
density of the black holes. Therefore, ensuring the independence of the bivariate truncated
data is the basis for us to get a meaningful statistical results from the BBH mass and
redshift observations \citep{1992ApJ...399..345E,1998astro.ph..8334E}.

In Figure~\ref{fig1}, there is a clear positive correlation between the observed mass and
redshift for both the pre-merger sample and the post-merger sample. This may be caused
by the observational selection effect \citep{2021PASA...38...56H, 2201.10258, 2023MNRAS.523.4539K}.
A GW detector has a limited sensitivity. It can sensitively record nearby GW events. But for
distant mergers, it can only detect those events involving relatively massive BHs.
Such an observational selection effect also needs to be corrected for before drawing
firm conclusions on the intrinsic distribution of BHs.

Therefore, in this section, we first
introduce the method to remove the dependence between mass and redshift and then
describe how to apply the Lynden-Bell's $C^{-}$ method to bivariate independent parameter
samples to obtain the BH mass function, redshift distribution and number density.

\subsection{The method of $\tau$ statistics}
\label{sec: 3.1}
The non-parametric $\tau$ statistic method \citep{1992ApJ...399..345E, 1998astro.ph..8334E} can
be applied to remove the bias caused by observational selection effects, i.e. the induced
correlation between $m$ and $z$. When this method is used in other fields such as
gamma-ray bursts and fast radio bursts, it is usually assumed that a power-law relation
exists between the luminosity and the redshift due to the observational selection
effects \citep{2020ApJ...890....8C, 2023JCAP...08..034B}. Similarly, in this study, we use
$m \propto (1+z)^{\gamma} $ to represent the biased dependence of mass on the redshift,
where $\gamma$ is a constant. Once the power-law index $\gamma$ is determined, we
then can correct the mass as $m' = m/(1+z)^{\gamma}$. In this way, we can get the independent
parameter pair of $m'$ and $z$, which means the BH distribution function can be expressed
as $\Psi(m',z) = \psi(m') \phi(z)$.

The value of $\gamma$ could be determined from the observational data. For a specific
$\gamma$, the observed data point of each black hole will change from $(z_i, m_i)$ to
the corrected point of $(z_i , m'_i)$. For the $i$th data point $(z_i, m'_i)$  in the BH sample,
we first define a data set $J_i$ as
\begin{equation}\label{Ji}
J_i = \{j|m'_j \geq m'_i, z_j\leq z_{i}^{\rm max}\},
\end{equation}
where $m'_i$ is the corrected mass of the $i$th BH and $z_{i}^{\rm max}$ is the maximum
redshift at which a BH with mass $m'_i$ can be detected by the detector. The number of BHs
contained in this region is denoted as $n_i$, and the number of BHs with redshift $z$ less than
or equal to $z_i$ in this region is designated as $R_i$. The $\tau$ test statistic is expressed as
\begin{equation}
\tau \equiv \frac{\sum_{i}(R_i - E_i)}{\sqrt{\sum_{i}{V_i}}},
\end{equation}
where $E_i = \frac{1+n_i}{2}$ and $V_i = \frac{(n_i- 1)^2}{12}$ are the expected mean value
and the variance of $R_i$, respectively.

According to the $\tau$ statistic test, if $R_i$ is uniformly distributed between 1 and $n_i$, then
the probability of $R_i \leq E_i$ and $R_i \geq E_i$ should be nearly equal so that we have $\tau=0$.
In this case, we could know that the distribution of mass and redshift are independent of each other.
It means that the assumed $\gamma$ value can correctly remove the bias introduced by the
observational selection effect. On the other hand, if $\tau$ does not equal zero, then we need
to adjust the value of $\gamma$. The above calculation process is repeated until $\tau = 0$ is
satisfied and the correct $\gamma$ value is determined.

During the calculations, the truncation line is a key ingredient
which is used to calculate $z_{i}^{\rm max}$. However, note that
the truncation line in Figure~\ref{fig1} is a complex function
which could not be described by a simple analytical equation.
Luckily, for our BH samples, the limit curve in the $z>1$ region
does not have any impact on the statistics of the non-parametric
method. Consequently, we are only concerned about the limit curve
in the $z<1$ segment. In this region, we notice that the limit
curve in fact could be well represented by a simple straight line
of $m=10z^{1.2}$ (i.e. the solid line in the figure). So, we use
this straight line as the the effective truncation limit to
perform our calculations.

\begin{figure*}[htbp]
     \centering
     \includegraphics[width=0.9\textwidth]{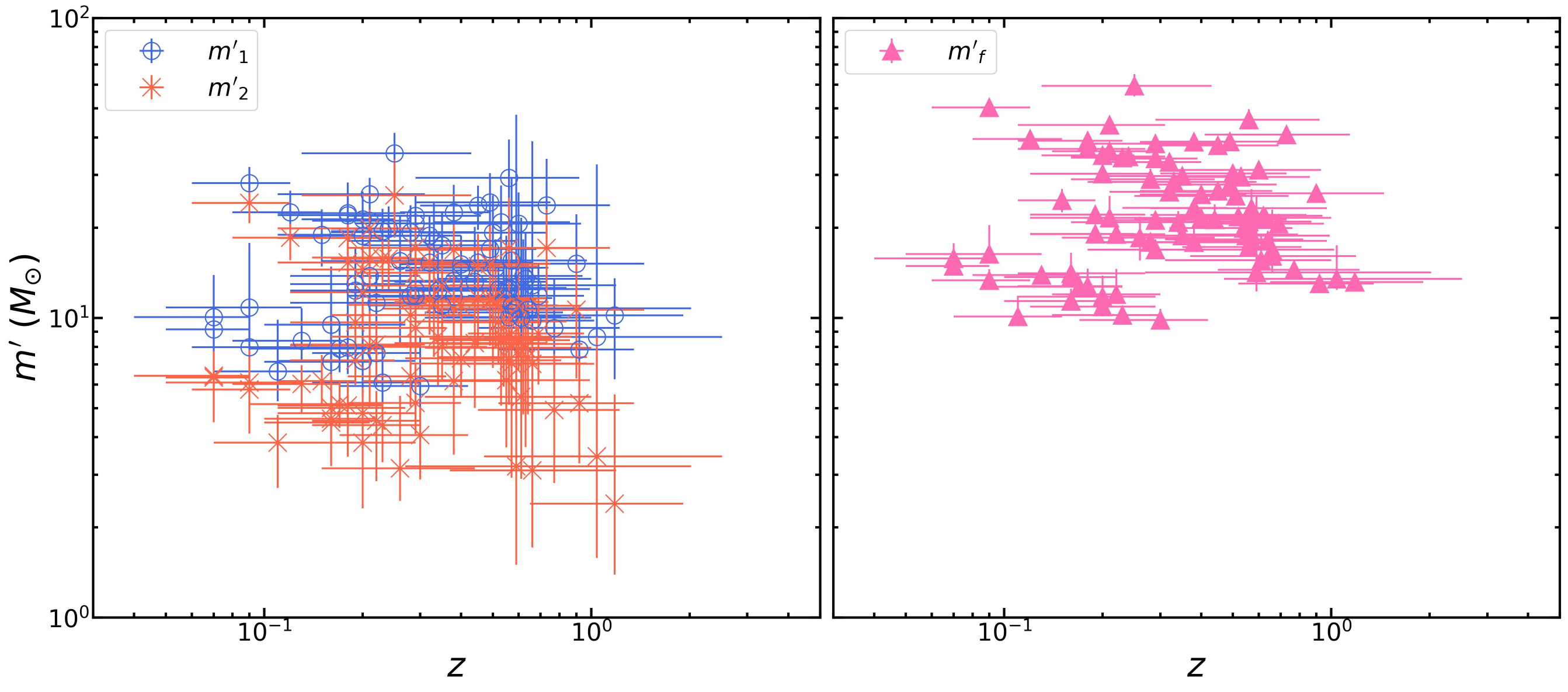}
     \caption{Distribution of the black hole masses and redshifts after correcting for
        the selection effect. Left panel: mass versus redshift for the pre-merger sample.
        The crosses and hollow circles represent the lighter ($m_2$) and heavier ($m_1$)
        black holes, respectively. Right panel: mass versus redshift for the post-merger sample.}
     \label{fig2}
\end{figure*}

Using this method of $\tau$ statistics, we finally get the best value as $\gamma = 2.72^{+0.31}_{-0.32}$ for
the pre-merger sample and $\gamma = 2.63^{+0.37}_{-0.38}$ for the post-merger sample. The observed
BH masses are then corrected by dividing them with $(1+z)^{\gamma}$. The distributions
of the BHs after correction are shown in Figure~\ref{fig2}. It can be seen that the masses of
the black holes no longer show any correlation with the redshifts.

\subsection{The Lynden-Bell's $C^{-}$ method }

The Lynden-Bell's $C^{-}$ method is an effective way to derive the bivariate distributions of
astronomical objects from the truncated data. In Equation~(\ref{Ji}), the number of BHs in
the region of $J_i$ is $n_i$. The Lynden-Bell's $C^{-}$ method does not include the $i$th BH
in the analysis, which means the BH number is smaller by 1, i.e. $N_i = n_i - 1$. This
is the reason that the method is called the ``$C^{-}$'' method \citep{1971MNRAS.155...95L}.
To carry out the calculations, we need to further define another set as
\begin{equation}\label{JI}
J^{\prime}_i = \{j|m'_{j} \geq m_{i}^{'\rm min}\ , z_j < z_i\},
\end{equation}
where $m_{i}^{'\rm min}$ is the limit mass at the redshift $z_i$. We denote the number of BHs
in $J^{\prime}_i$ as $M_i$.

According to the Lynden-Bell's $C^{-}$ method, the cumulative mass function can then be
calculated as \citep{1971MNRAS.155...95L, 1992ApJ...399..345E}
\begin{equation}\label{massFunction}
\psi(m'_{i}) = \prod\limits_{j>i}(1+\frac{1}{N_j}),
\end{equation}
where $j>i$ means that the operation applies to all BHs whose mass $m'_{j}$ is larger than $m'_{i}$.
Similarly, the cumulative redshift distribution $\phi(z)$ can be expressed as
\begin{equation}
\label{redshiftFunction}
\phi(z_i) = \prod\limits_{j<i}(1+\frac{1}{M_j}),
\end{equation}
where $j<i$ means that the operation applies to all BHs whose redshift $z_j$ is less than $z_i$.

The number density of BHs can be written as
\begin{equation}\label{formationrate}
\rho_{BH}(z) = \frac{d\phi(z)}{dz}(1+z)(\frac{dV(z)}{dz})^{-1},
\end{equation}
where the term of $(1+z)$ results from the cosmological time dilation and $dV(z)/dz$ is the
differential comoving volume which can be further expressed as \citep{2019JHEAp..24....1K}
\begin{equation}\label{comovingvolume}
\frac{dV(z)}{dz}=\frac{c}{H_0}\frac{4\pi{d^2_{l}(z)}}{(1+z)^2}\frac{1}{\sqrt{{\Omega}_{\Lambda}+\Omega_{\rm m}(1+z)^3}}.
\end{equation}
Note that the comoving volume at a redshift of $z$ is $V = 4 \pi {D_M^3} / 3$,
where the comoving distance is $D_M = d_l / (1+z)$ (\citealt{1999astro.ph..5116H}).

\subsection{Test of the method}

The GW observation can be approximately
modelled as a threshold process of the observed SNR, which itself
depends not only on the GW parameters but also on the specific
noise \citep{2018ApJ...863L..41F, 2019MNRAS.486.1086M,
2024ApJ...962..169E, 2024CQGra..41l5002G}. So the observed
distribution cannot be simply modelled as a truncated version of
the intrinsic one without introducing a significant bias.

\begin{figure*}[htbp]
     \centering
     \includegraphics[width=0.9\textwidth]{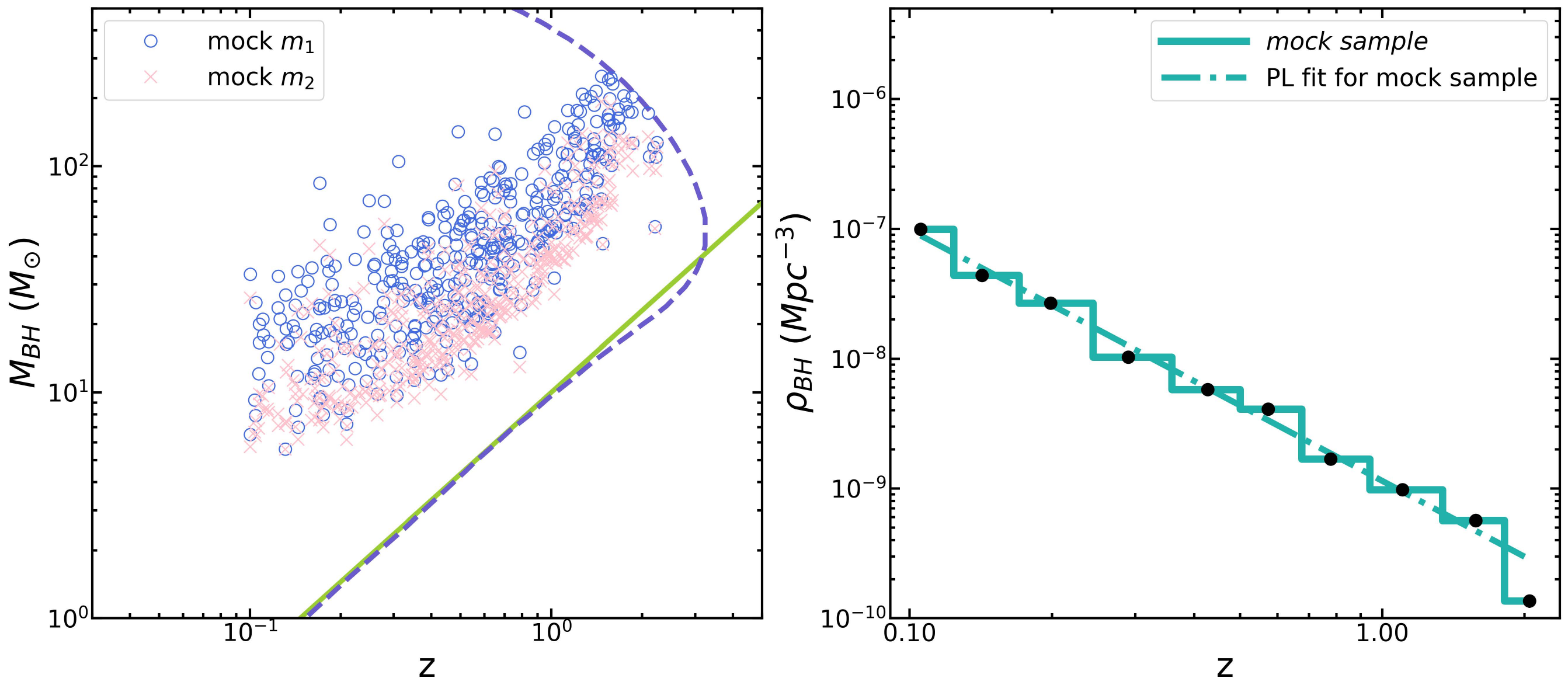}
     \caption{ Testing the non-parametric method by using completely
     simulated BBH events.
     Left panel: distribution of the mock BHs on the mass-redshift plane.
     The crosses and hollow circles represent the lighter (mock $m_2$) and
     heavier (mock $m_1$) black holes, respectively. Right panel: number density
     versus redshift derived from the mock BHs. The dash-dotted
     line represents the best-fit power-law function.}
     \label{fig3}
\end{figure*}

To assess the credibility of our method
and the reliability of the results, we have performed simulation
tests by using completely simulated samples. For this purpose, we
first assume an intrinsic number density of $\rho(z) \propto
z^{-2.0}$. The number density is then combined with the observed
mass function (see results in the next section) to generate a mock
BBH sample of 1000 events through Monte Carlo simulations. The
distribution of the mock BBHs on the mass-redshift plane is shown
in the left panel of Figure~\ref{fig3}. The non-parametric method
described above is then used to analyze the mock, trying to
reconstruct the underlined number density. The reconstructed
results are shown in the right panel of Figure~\ref{fig3}. In this
case, the best-fit power-law function for the number density is
$\rho(z) \propto z^{-1.94 \pm 0.15}$, which is well consistent
with the pre-assumed function of $\rho(z) \propto z^{-2.0}$.

In fact, we have also assumed the number
density as $\rho(z) \propto z^{-1.8}$ and $\rho(z) \propto
z^{-2.2}$ in two other simulations. The recovered number density
function is $\rho(z) \propto z^{-1.83 \pm 0.18 }$ and $\rho(z)
\propto z^{-2.22 \pm 0.13}$, respectively. In all these cases, we
see that the non-parametric method can satisfactorily re-construct
the pre-assumed number density function, which means the method is
effective and reliable.

\section{results}
\label{sec: RESULTS}

We have applied the unbiased non-parametric $\tau$ statistics
method on the pre-merger and post-merger BH samples to correct for
the dependence between mass and redshift of BHs induced by the
observational selection effects. The Lynden-Bell's $C^{-}$ method is
then used to derive the intrinsic mass function and redshift
distribution of BHs based on the two samples. Additionally, the
number density of BHs in the Universe is inferred by considering
the redshift information of the BHs. Here we present our numerical
results as follows.

\subsection{Mass distribution function}

\begin{figure}[htbp]
     \centering
     \includegraphics[width=0.45\textwidth]{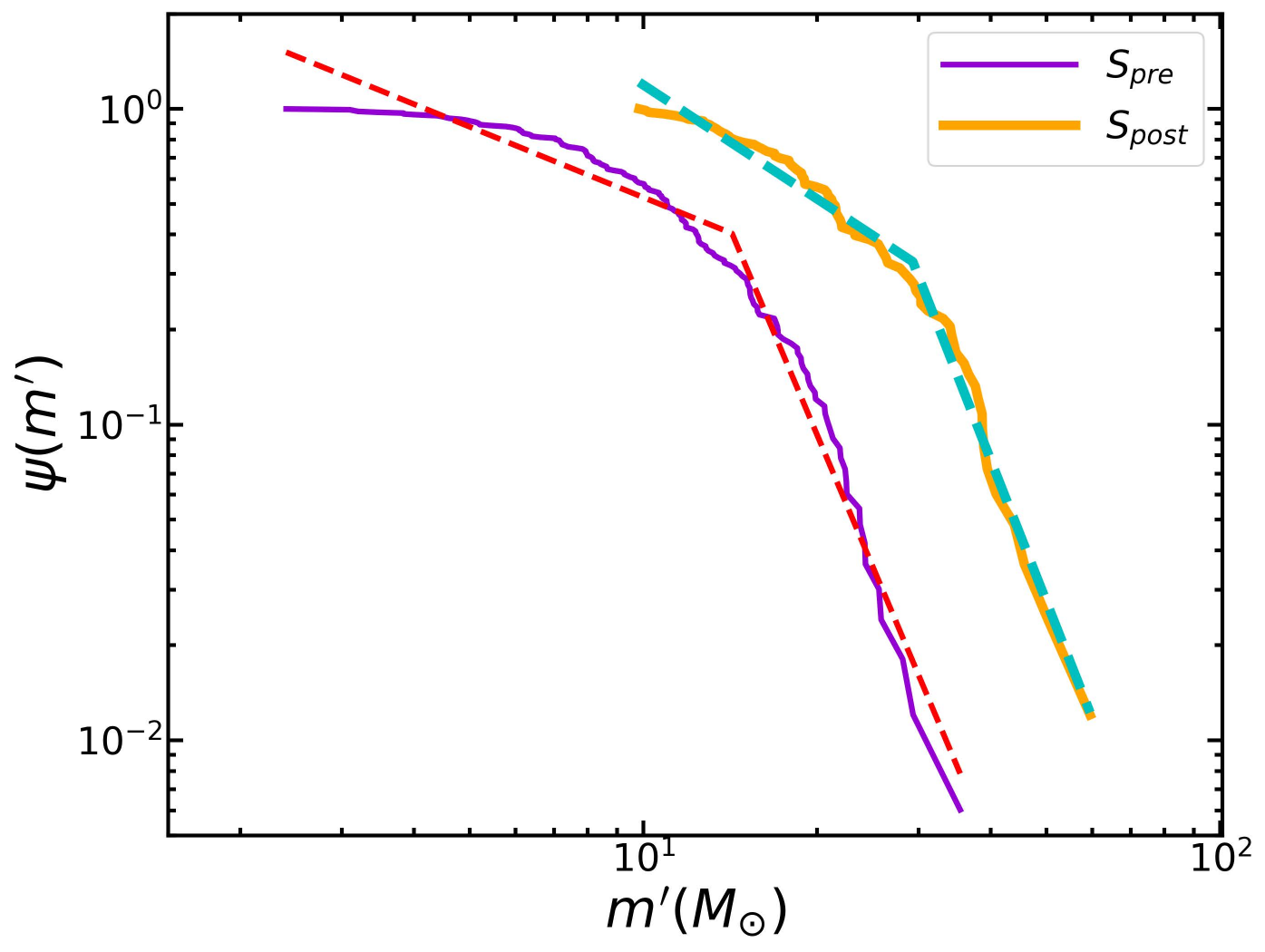}
     \caption{Mass function of BHs. The thin solid line represents the pre-merger
       sample and the thick solid line represents the post-merger sample. Both samples
      are normalized by the maximum value of the function, respectively. The thin and
      thick dashed lines represent the best fitting results by using a broken power-law
      function correspondingly.}
     \label{fig4}
\end{figure}

The mass distribution function of BHs can be calculated by using Equation~(\ref{massFunction}).
The results derived from the pre-merger sample and the post-merger sample are shown in
Figure~\ref{fig4}. We see that for both samples, the mass function shows a decreasing trend
with the increasing mass. However, it is obvious that the mass distribution of the pre-merger
sample is different from that of the post-merger sample. In the former case, the mass is mainly
distributed between 2 $M_{\odot}$ and 40 $M_{\odot}$; while it is in a range
of 10 $M_{\odot}$ -- 60 $M_{\odot}$ in the latter case. It clearly shows that the post-merger
BHs are significantly more massive.

We have used a simple broken power-law function to fit the mass
distribution function, i.e.
\begin{equation}
\label{broken power law}
f(m) \propto \left\{
\begin{array}{ll}
        (m/m_b)^{\alpha}, \ \
        & m \leq m_b, \\
        (m/m_b)^{\beta}, \ \
        & m > m_b,
\end{array}
\right.
\end{equation}
where $m_b$ is the mass at the broken position, $\alpha$ and $\beta$ are the two power-law
indices characterizing the steepness of the mass function before and after the broken.
The best fitting results of the pre-merger sample are $\alpha = -0.74 \pm 0.21$,
$\beta = -4.32 \pm 0.58$, and $m'_b = 14.24 \pm 1.06 M_{\odot}$. For the post-merger sample,
the best fitting results are $\alpha = -1.2 \pm 0.42$, $\beta= -4.65 \pm 1.15$,
and $m'_b = 29.31 \pm 3.61 M_{\odot}$. The results are shown by the dashed lines in Figure~\ref{fig4}.
We see that the broken mass of $m'_b$ of the post-merger sample is significantly larger than that
of the pre-merger sample. This is easy to understand since the BHs in the post-merger sample are
generally more massive. It is interesting to note that the power-law index in the high mass segment,
i.e. $\beta$, is generally consistent with each other for the two samples, which means that the mass
function of BHs has a steep index of $\beta \approx -4$ -- $-5$. On the other hand, the index of
$\alpha$ is clearly different for the two samples. The reason may be that there are too few less
massive BHs in the post-merger sample.

Different structures can be identified in
the mass distribution of BBHs. For example, it can be modelled
with a broken power-law function, a power-law function with a
sharp high-mass cut-off, or a Gaussian feature superposed on a
power-law function \citep{2019ApJ...882L..24A, 2021ApJ...913L...7A}.
Our sample is still relatively small,
which only contains 83 BBH merger events.
Although our results strongly support the broken power-law model
for the mass function, we would like to point out that other
models such as a power-law function plus a Gaussian component or a
more complex model still cannot be ruled out due to the very
limited sample size. A significantly expanded sample available in
the future would be helpful in this aspect.

\cite{2021ApJ...913L...7A} noticed that
the distribution of the primary mass of the observed BBH events
can be described by a broken power-law function, with a broken
mass of $39.7^{+20.3}_{-9.1} M_{\odot}$. It could also be modelled
by a Gaussian component peaking at $33^{+4.0}_{-5.6} M_{\odot}$,
superposed on a power-law function. In our study, as shown in
Figure~\ref{fig4}, the mass function of the pre-merger BHs is
better described by a broken power-law function, with a broken
mass of $m'_b = 14.24 \pm 1.06 M_{\odot}$. Note that $m'_b$ here
is the de-evolved mass (after correcting for the redshift
evolution). The original broken mass is actually $m_b =
m'_b(1+z)^{2.72}$ (see Section~\ref{sec: 3.1}). Taking the average redshift
of all the BBHs as $z=0.38$, we get $m_b = 34.2 \pm 3.8
M_{\odot}$. This broken mass is interestingly comparable to the
characteristic mass reported by \cite{2021ApJ...913L...7A}.


\subsection{Redshift distribution and number density of BHs}
\label{Redshift distribution and number density of BHs}
\begin{figure}[htbp]
     \centering
     \includegraphics[width=0.46\textwidth]{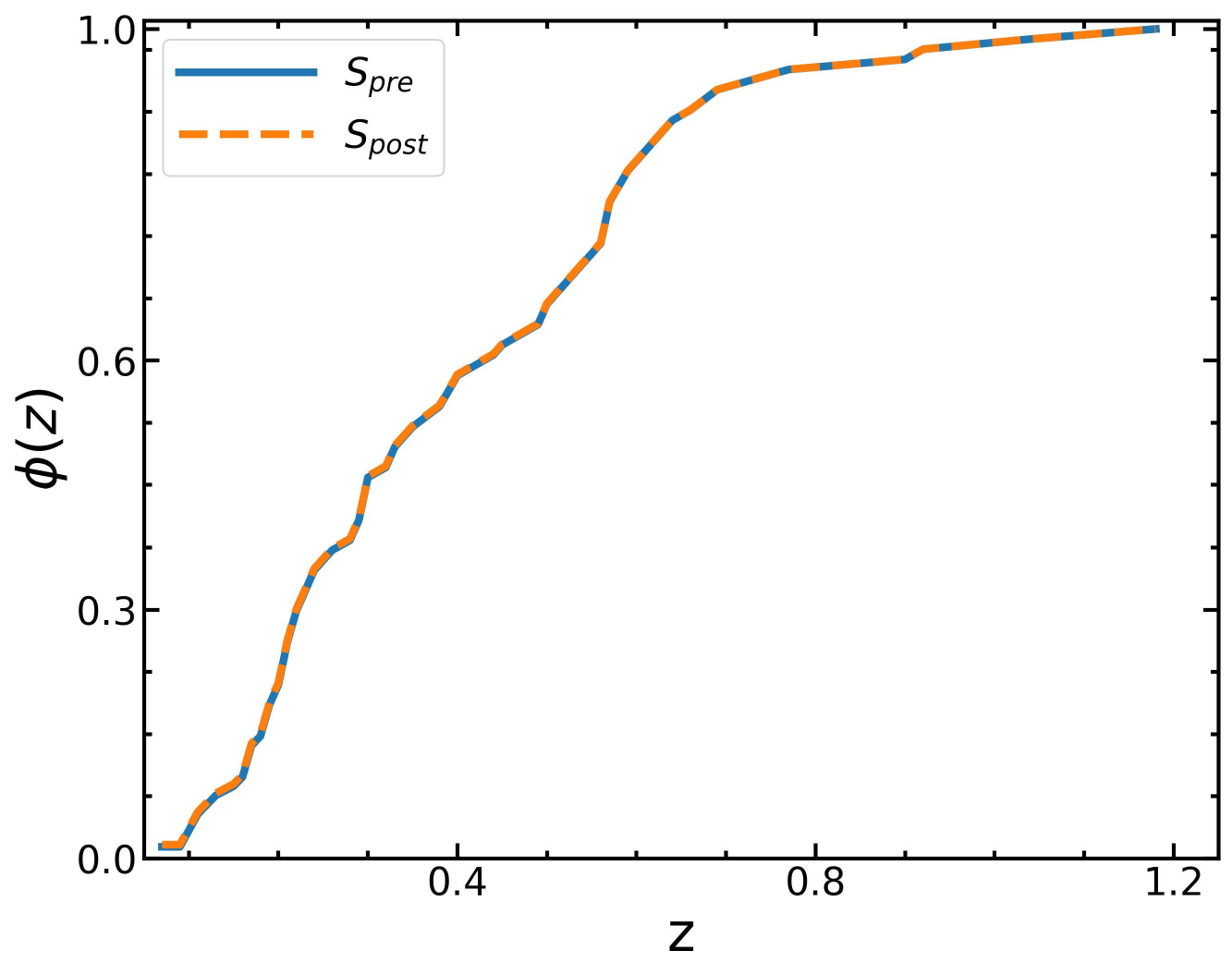}
     \caption{Cumulative redshift distribution of BHs. The solid line represents the pre-merger
         sample and the dashed line represents the post-merger sample. The two curves are
         essentially overlapped with each other. }
     \label{fig5}
\end{figure}

The cumulative redshift distribution of the two BH samples can be calculated by using
Equation~(\ref{redshiftFunction}). The results are shown in Figure~\ref{fig5}. We see that the
cumulative distribution is almost identical for the two samples. This is easy to understand, since
each post-merger BH corresponds to two pre-merger BHs at a particular redshift.  Also, it is
interesting to note that the cumulative distribution increases rapidly at two redshifts, i.e.
$z \sim 0.2$ and $z \sim 0.55$. It indicates that there are more BHs at these two distances.

\begin{figure}[htbp]
     \centering
     \includegraphics[width=0.45\textwidth]{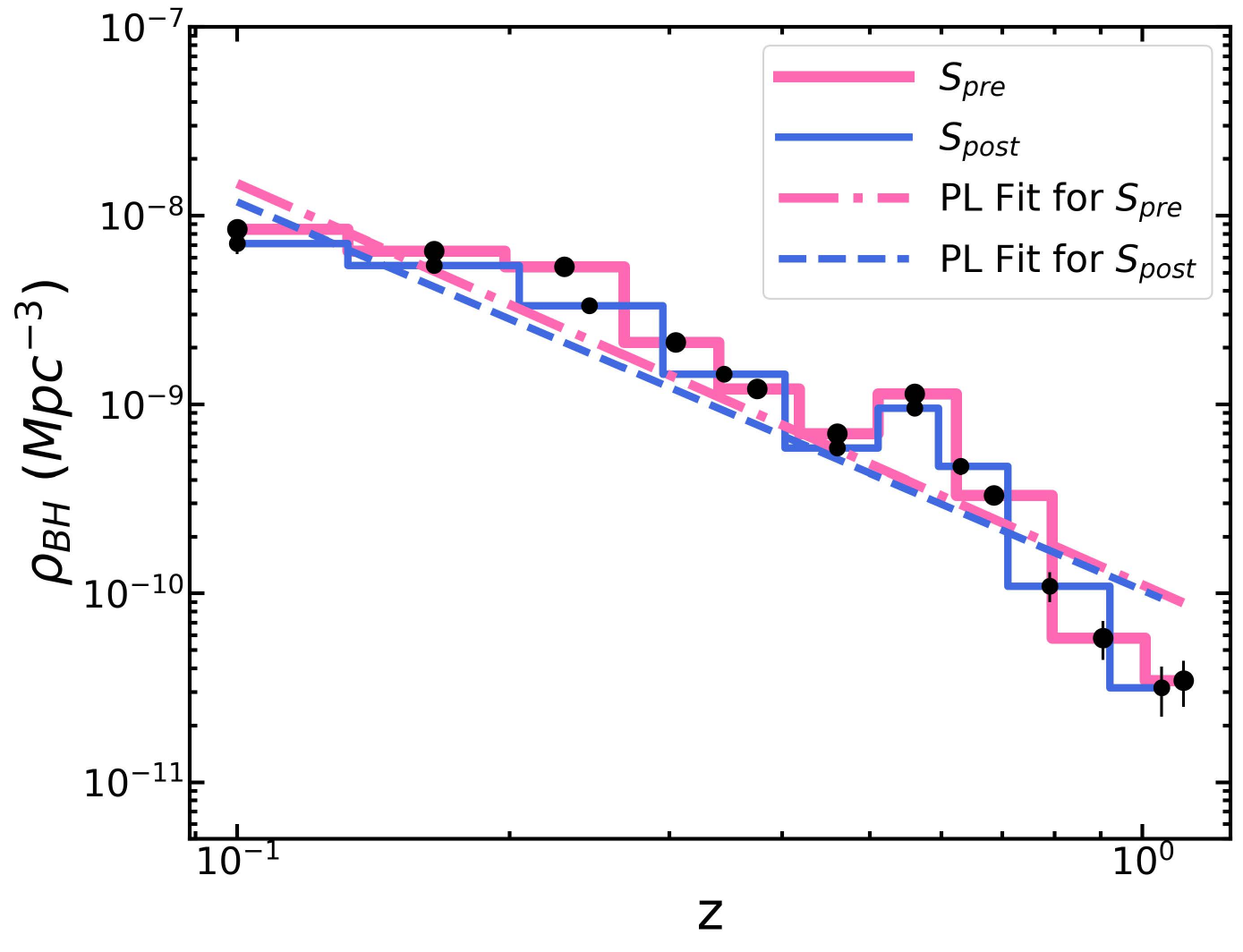}
     \caption{The number density of BHs versus redshift. The thick step line represents the pre-merger
         sample and the thin step line represents the post-merger sample. The dash-dotted line and the dashed
        line represent the best fitting results of the pre-merger sample and the post-merger sample, respectively,
        by using a simple power-law function. }
     \label{fig6}
\end{figure}

After obtaining the redshift distribution, we have further calculated the number density of black
holes by using Equation~(\ref{formationrate}). The number density is plot versus redshift in
Figure~\ref{fig6} for the two BH samples. We see that for both BH samples, the number density
decreases steadily as the redshift increases. The behavior can be well fited by a power-law function
in the form of $\rho(z) \propto z^B$, where $B$ is the power-law index.  For the pre-merger sample,
the best-fit value is $B\sim-2.12 \pm 0.04$ . For the post-merger sample, we have $B\sim-2.06 \pm 0.04$. Interestingly,
these two indices are consistent with each other, suggesting that the number density of BHs in the
two samples has a similar redshift dependence. Furthermore, from Figure~\ref{fig6}, we can notice
that the number density has a slight excess at $z \sim$ 0.2 -- 0.3 and $z \sim$ 0.5 -- 0.6, which is
exactly the reason that leads to the two rapidly increasing episodes illustrated in Figure~\ref{fig5}.

\subsection{Further test of the method by expanding the sample}

\begin{figure*}[htbp]
     \centering
     \includegraphics[width=0.9\textwidth]{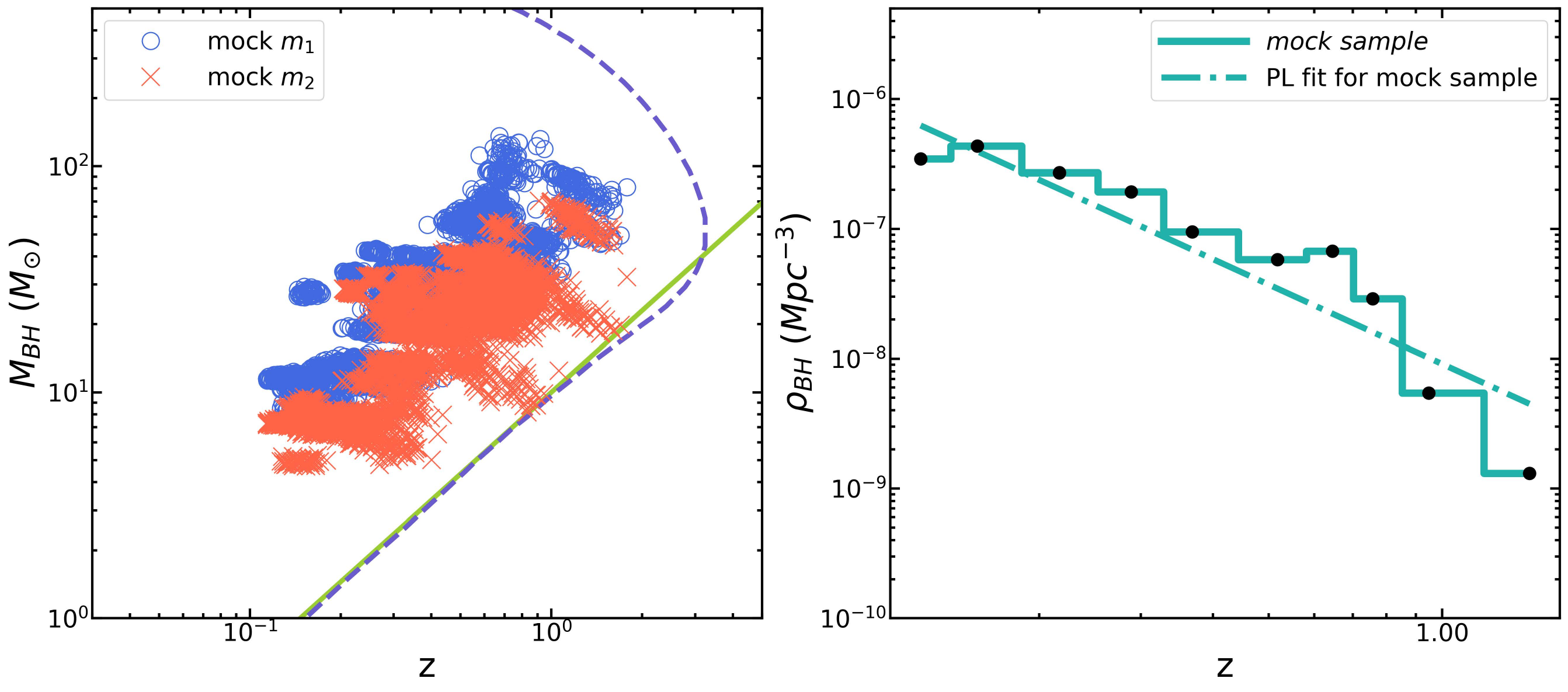}
     \caption{Testing the non-parametric
     method by expanding the observed BBH sample to a much larger
     mock BBH sample. Left panel: distribution of the mock BHs on the mass-redshift plane.
     The crosses and hollow circles represent the lighter (mock $m_2$) and
     heavier (mock $m_1$) black holes, respectively. Right panel: number density
     versus redshift derived from the mock BHs. The solid step line represents the
     results derived directly by using our method. The dash-dotted
     line represents the best-fit power-law function.}
     \label{fig7}
\end{figure*}

Following \cite{2020ApJ...891L..31F}, we
have generated a much larger mock BBH sample to further examine
the credibility of our method. For each merger event in the
observed sample (which includes 83 BBHs), we randomly take 50 SNR
values from a normal distribution centered at the true SNR to
generate the mock SNRs. The resulting sample contains 4150 mock
BBHs. Similar to \cite{2020ApJ...891L..31F}, the masses and
redshifts of the mock BBHs are randomly evaluated. The
distribution of the mock BBH sample on the mass-redshift plane is
shown in the left panel of Figure~\ref{fig7}. The non-parametric
method is then applied to analyze the mock sample. The number
density derived from the mock sample is shown in the right panel
of Figure~\ref{fig7}. The best-fit power-law index of the plot is
$\sim -2.03\pm0.05$, which is consistent with the result ($B \sim
-2.12 \pm 0.04$, see Section~\ref{Redshift distribution and number
density of BHs}) derived from the real observed sample. From this
test, it is quite clear that the non-parametric method is very
effective and credible in analyzing the BBH GW event data.

As mentioned earlier in Section~\ref{sec: 3.1}, a
pre-determined truncation line is necessary in the practice of the
non-parametric method. The truncation line is affected by many
factors concerning the instrumentation and observation, thus is
always ambiguous to some extent. Luckily, it would not affect the
final result significantly as long as most of the observational
data points are above the adopted truncation line on the
$M_{BH}$-$z$ plane. In fact, the influence of the truncation line
has been examined in many similar applications of the
non-parametric method in other field such as gamma-ray bursts
\citep{2021ApJ...914L..40D, 2022MNRAS.513.1078D}. It is found that
the final result is not sensitive to the set up of the truncation
line.

\section{Conclusions and Discussion}
\label{sec: Conclusion}

Ninety credible GW events are released in the GWTC catalogue by
the LVK collaboration, among which 83 cases are confirmed binary
black hole mergers. In this study, we have carried out a
statistical analysis on the masses and redshifts of these BHs. Two
samples consisted of the 166 pre-merger BHs and the 83 post-merger
BHs are considered separately, with the former sample representing
BHs that once exist in the Universe and the latter representing
newly born BHs. The non-parametric $\tau$ statistic method is
adopted to remove the bias caused by observational selection
effect. The sensitivity curve of upgraded
(A+) Advanced LIGO is used to calculate the truncation line. The
derived dependence between redshift and mass are $m \propto
(1+z)^{2.72}$ for the pre-merger sample and $m \propto
(1+z)^{2.63}$ for the post-merger sample. After correcting for the
selection effect, the mass and redshift become two independent
parameters for the two samples, based on which meaningful
statistical analysis can be carried out. Using the Lynden-Bell's
$C^{-}$ method, it is found that the mass distribution can be
expressed as a broken power-law function for both samples. In the
case of the pre-merger sample, the power-law indices are $\alpha =
-0.74 \pm 0.21$ and $\beta = -4.32 \pm 0.58$ in the smaller mass
region and the higher mass region, respectively, with the break
occurring at $m'_b = 14.24 \pm 1.06 M_{\odot}$. For the
post-merger sample, the indices are $\alpha = -1.2 \pm 0.42$ and
$\beta= -4.65 \pm 1.15$ correspondingly, with the broken mass
being $m'_b = 29.31 \pm 3.61 M_{\odot}$. We notice that the
power-law index $\beta$ in the higher mass region is essentially
identical for the two samples. Additionally, the number density of
BHs is derived as $\rho(z) \propto z^{-2.06}$ -- $z^{-2.12}$ from
the two samples.

The merger rate of binary black hole systems has been explored by many
groups \citep{2019MNRAS.485..889S, 2020PhRvD.102l3016A, 2021ApJ...913L...7A, 2022ApJ...931...17V, 2023ApJ...953..153H, 2023MNRAS.522.5546F,2023MNRAS.523.5565C,2024ApJ...961....8F},
with different conclusions being reached. For example, \cite{2020ApJ...890....8C} studied the merger process
of black holes in ultra low-mass dwarf galaxies. Although their merger rate is high enough to
account for the gravitational wave events detected by LIGO/Virgo, a special feature of their
results is that the rate increases with the increasing redshift under various time delay assumptions.
On the other hand, \cite{2024ApJ...972..157V} argued that if BBHs were hosted by a sample of
galaxies purely weighted by their stellar mass, then the BBH number density would roughly decrease
with the increasing redshift as $(1+z)^{-0.64}$. The different outcomes on BBH rates may result from
different assumptions and different samples being used. It also reflects the complexity of this difficult
problem. In our study, we have used a non-parametric method in which as fewer assumptions are made
as possible. Additionally, our study is based on the unified GW events detected by the LVK collaboration.
So, it is a beneficial attempt on this issue.

According to our study, the number
density of BHs scales with the redshift as $\rho(z) \propto
z^{-2.06}$ -- $z^{-2.12}$, which indicates that the number density
decreases with the increasing redshift. It is interesting to note
that \cite{2023PhRvX..13a1048A} found that the rate of BBH mergers
is proportional to $(1+z)^{{2.9}^{+1.7}_{-1.8}}$, i.e. increasing
with redshift. Since the bases of the two power-law expressions
are different (i.e. $z$ versus $1+z$), a direct comparison of the
two power-law indices is not available. Here we would like to
point out that Abbott et al.'s expression actually corresponds to
the term of $\frac{d\phi(z)}{dz}$ in our
Equation~\ref{formationrate}. It differs with $\rho(z)$ by a
factor of $(1+z)(\frac{dV(z)}{dz})^{-1}$, which is a complicated
term involving the redshift (see Equations~(\ref{formationrate})
and~(\ref{comovingvolume})). Additionally, our expression refers
to the transformed (and de-correlated) samples, which further
explains the huge difference between our index and the index
reported by \cite{2023PhRvX..13a1048A}.

It deserves mentioning that the BHs in our post-merger samples are universally produced via the merging
of two smaller BHs. However, the origin of the BHs in the pre-merger sample is uncertain. Theoretically,
they could come from direct collapse of old massive stars, growth of less massive BHs through accretion,
or merging of two BHs. The similarity of some features between the two samples, such as the consistent
power-law index $\beta$ of the mass distribution function in the higher mass region, and the almost equal
power-law index $B$ of the number density, could potentially provide interesting information on the
origin of the pre-merger BHs. It points to the possibility that the majority of them, especially those at the
higher mass end, should also be of merger origin.
However, note that $S_{pre}$ and $S_{post}$ are not independent samples
because they are filtered in the same way. Especially, the masses of $S_{post}$ are connected
to the masses of $S_{pre}$. This may also be part of the reason that the results of
post-merger sample and pre-merger sample are similar. Meanwhile, the number of BHs in the two samples is still very
limited currently. The fourth observing run (O4) of LVK is still in progress. When it ends in February 2025, the total
number of detected GW events is expected to exceed 200. A much larger catalogue of BBHs will be available,
which will be helpful for us to further explore the origin and distribution of BHs in the Universe.

\section{Acknowledgements}

We thank the anonymous referee for useful comments and suggestions
that led to an overall improvement of this study. We are grateful
to Yi-Han Iris Yin for helpful discussions. This study was
supported by the National Natural Science Foundation of China
(Grant Nos. 12233002, U2031118), by the National SKA Program of
China No. 2020SKA0120300, by the National Key R\&D Program of
China (2021YFA0718500). LXJ acknowledges the support by Shandong
Provincial Natural Science Foundation (Grant No. ZR2023MA049). YFH
also acknowledges the support from the Xinjiang Tianchi Program.

\bibliography{refs}

\clearpage
\centering
\begin{longtable}{p{3.2cm}<{\centering}p{2cm}<{\centering}p{2cm}<{\centering}p{2cm}<{\centering}p{2cm}<{\centering}p{2.2cm}<{\centering}p{1.5cm}}
    \caption{Key parameters of the black holes in the BBH GW events.}
    \label{tab:1}\\
    \hline%
    \hline%
    $\rm Name$ & $m_{1}$ ($M_{\odot}$) & $m_{2}$ ($M_{\odot}$) & $m_{f}$ ($M_{\odot}$) & $\rm redshift$ & $ \rm FAR (year^{-1}) $ & $p_{astro} $\\
    \hline%
    \endfirsthead
    \multicolumn{6}{c}{\textbf{Table 1}-\textit{continued}} \\
    \hline
    \hline
    $\rm Name$ & $m_{1}$ ($M_{\odot}$) & $m_{2}$ ($M_{\odot}$) & $m_{f}$ ($M_{\odot}$) & $ \rm redshift  $ & $ \rm FAR (year^{-1})$ & $p_{astro} $ \\
    \hline
    \endhead
    \hline
    \endfoot
    \hline
    \hline
    \endlastfoot
    $\rm GW150914$ & $35.6^{+4.7}_{-3.1}$ & $30.6^{+3.0}_{-4.4}$ & $63.1^{+3.4}_{-3.0}$ & $0.09^{+0.03}_{-0.03}$ & $1.00E-07$ & 1.00  \\
    $\rm GW151012$ & $23.2^{+14.9}_{-5.5}$ & $13.6^{+4.1}_{-4.8}$ & $35.6^{+10.8}_{-3.8}$ & $0.21^{+0.09}_{-0.09}$ & $7.92E-03$ & 1.00  \\
    $\rm GW151226$ & $13.7^{+8.8}_{-3.2}$ & $7.7^{+2.2}_{-2.5}$ & $20.5^{+6.4}_{-1.5}$ & $0.09^{+0.04}_{-0.04}$ & $1.00E-07$ & 1.00  \\
    $\rm GW170104$ & $30.8^{+7.3}_{-5.6}$ & $20.0^{+4.9}_{-4.6}$ & $48.9^{+5.1}_{-4.0}$ & $0.20^{+0.08}_{-0.08}$ & $1.00E-07$ & 1.00  \\
    $\rm GW170608$ & $11.0^{+5.5}_{-1.7}$ & $7.6^{+1.4}_{-2.2}$ & $17.8^{+3.4}_{-0.7}$ & $0.07^{+0.02}_{-0.02}$ & $1.00E-07$ & 1.00  \\
    $\rm GW170729$ & $50.2^{+16.2}_{-10.2}$ & $34.0^{+9.1}_{-10.1}$ & $79.5^{+14.7}_{-10.2}$ & $0.49^{+0.19}_{-0.21}$ & $2.00E-02$ & 0.98  \\
    $\rm GW170809$ & $35.0^{+8.3}_{-5.9}$ & $23.8^{+5.1}_{-5.2}$ & $56.3^{+5.2}_{-3.8}$ & $0.20^{+0.05}_{-0.07}$ & $1.00E-07$ & 1.00  \\
    $\rm GW170814$ & $30.6^{+5.6}_{-3.0}$ & $25.2^{+2.8}_{-4.0}$ & $53.2^{+3.2}_{-2.4}$ & $0.12^{+0.03}_{-0.04}$ & $1.00E-07$ & 1.00  \\
    $\rm GW170818$ & $35.4^{+7.5}_{-4.7}$ & $26.7^{+4.3}_{-5.2}$ & $59.4^{+4.9}_{-3.8}$ & $0.21^{+0.07}_{-0.07}$ & $4.20E-05$ & 1.00  \\
    $\rm GW170823$ & $39.5^{+11.2}_{-6.7}$ & $29.0^{+6.7}_{-7.8}$ & $65.4^{+10.1}_{-7.4}$ & $0.35^{+0.15}_{-0.15}$ & $1.00E-07$ & 1.00  \\
    $\rm GW190403\_051519$ & $85.0^{+27.8}_{-33.0}$ & $20.0^{+26.3}_{-8.4}$ & $102.2^{+26.3}_{-24.3}$ & $1.18^{+0.73}_{-0.53}$ & $7.70E+00$ & 0.61  \\
    $\rm GW190408\_181802$ & $24.8^{+5.4}_{-3.5}$ & $18.5^{+3.3}_{-4.0}$ & $41.4^{+3.9}_{-2.9}$ & $0.29^{+0.07}_{-0.11}$ & $1.00E-05$ & 1.00  \\
    $\rm GW190412$ & $27.7^{+6.0}_{-6.0}$ & $9.0^{+2.0}_{-1.4}$ & $35.6^{+4.8}_{-4.5}$ & $0.15^{+0.04}_{-0.04}$ & $1.00E-05$ & 1.00  \\
    $\rm GW190413\_052954$ & $33.7^{+10.4}_{-6.4}$ & $24.2^{+6.5}_{-7.0}$ & $55.5^{+10.1}_{-7.3}$ & $0.56^{+0.25}_{-0.21}$ & $8.20E-01$ & 0.93  \\
    $\rm GW190413\_134308$ & $51.3^{+16.6}_{-12.6}$ & $30.4^{+11.7}_{-12.7}$ & $78.0^{+16.1}_{-11.5}$ & $0.62^{+0.32}_{-0.26}$ & $1.80E-01$ & 0.99  \\
    $\rm GW190421\_213856$ & $42.0^{+10.1}_{-7.4}$ & $32.0^{+8.3}_{-9.8}$ & $70.5^{+12.4}_{-9.0}$ & $0.45^{+0.21}_{-0.19}$ & $2.80E-03$ & 1.00  \\
    $\rm GW190426\_190642$ & $105.5^{+45.3}_{-24.1}$ & $76.0^{+26.2}_{-36.5}$ & $172.9^{+37.7}_{-33.6}$ & $0.73^{+0.41}_{-0.32}$ & $4.10E+00$ & 0.75  \\
    $\rm GW190503\_185404$ & $41.3^{+10.3}_{-7.7}$ & $28.3^{+7.5}_{-9.2}$ & $66.5^{+9.4}_{-7.9}$ & $0.29^{+0.10}_{-0.10}$ & $1.00E-05$ & 1.00  \\
    $\rm GW190512\_180714$ & $23.2^{+5.6}_{-5.6}$ & $12.5^{+3.5}_{-2.6}$ & $34.3^{+4.1}_{-3.4}$ & $0.28^{+0.08}_{-0.10}$ & $7.92E-03$ & 1.00  \\
    $\rm GW190513\_205428$ & $36.0^{+10.6}_{-9.7}$ & $18.3^{+7.4}_{-4.7}$ & $52.1^{+8.8}_{-6.6}$ & $0.40^{+0.14}_{-0.13}$ & $1.30E-05$ & 1.00  \\
    $\rm GW190514\_065416$ & $40.9^{+17.3}_{-9.3}$ & $28.4^{+10.0}_{-10.1}$ & $66.4^{+19.0}_{-11.5}$ & $0.64^{+0.33}_{-0.30}$ & $2.80E+00$ & 0.76  \\
    $\rm GW190517\_055101$ & $39.2^{+13.9}_{-9.2}$ & $24.0^{+7.4}_{-7.9}$ & $60.1^{+9.9}_{-9.4}$ & $0.33^{+0.26}_{-0.15}$ & $3.50E-04$ & 1.00  \\
    $\rm GW190519\_153544$ & $65.1^{+10.8}_{-11.0}$ & $40.8^{+11.5}_{-12.7}$ & $100.0^{+13.0}_{-12.9}$ & $0.45^{+0.24}_{-0.15}$ & $1.00E-05$ & 1.00  \\
    $\rm GW190521$ & $98.4^{+33.6}_{-21.7}$ & $57.2^{+27.1}_{-30.1}$ & $147.4^{+40.0}_{-16.0}$ & $0.56^{+0.36}_{-0.27}$ & $1.30E-03$ & 1.00  \\
    $\rm GW190521\_074359$ & $43.4^{+5.8}_{-5.5}$ & $33.4^{+5.2}_{-6.8}$ & $72.6^{+6.5}_{-5.4}$ & $0.21^{+0.10}_{-0.10}$ & $1.00E-05$ & 1.00  \\
    $\rm GW190527\_092055$ & $35.6^{+18.7}_{-8.0}$ & $22.2^{+9.0}_{-8.7}$ & $55.5^{+17.9}_{-8.5}$ & $0.44^{+0.29}_{-0.19}$ & $1.00E-07$ & 1.00  \\
    $\rm GW190602\_175927$ & $71.8^{+18.1}_{-14.6}$ & $44.8^{+15.5}_{-19.6}$ & $110.5^{+17.9}_{-13.9}$ & $0.49^{+0.26}_{-0.20}$ & $1.00E-05$ & 1.00  \\
    $\rm GW190620\_030421$ & $58.0^{+19.2}_{-13.3}$ & $35.0^{+13.1}_{-14.5}$ & $88.0^{+17.2}_{-12.4}$ & $0.50^{+0.23}_{-0.20}$ & $1.10E-02$ & 0.99  \\
    $\rm GW190630\_185205$ & $35.1^{+6.5}_{-5.5}$ & $24.0^{+5.5}_{-5.2}$ & $56.6^{+4.4}_{-4.5}$ & $0.18^{+0.09}_{-0.07}$ & $1.00E-05$ & 1.00  \\
    $\rm GW190701\_203306$ & $54.1^{+12.6}_{-8.0}$ & $40.5^{+8.7}_{-12.1}$ & $90.2^{+11.2}_{-8.9}$ & $0.38^{+0.11}_{-0.12}$ & $5.70E-03$ & 1.00  \\
    $\rm GW190706\_222641$ & $74.0^{+20.1}_{-16.9}$ & $39.4^{+18.4}_{-15.4}$ & $107.3^{+25.2}_{-15.9}$ & $0.60^{+0.33}_{-0.29}$ & $5.00E-05$ & 1.00  \\
    $\rm GW190707\_093326$ & $12.1^{+2.6}_{-2.0}$ & $7.9^{+1.6}_{-1.3}$ & $19.2^{+1.7}_{-1.2}$ & $0.17^{+0.06}_{-0.08}$ & $1.00E-05$ & 1.00  \\
    $\rm GW190708\_232457$ & $19.8^{+4.3}_{-4.3}$ & $11.6^{+3.1}_{-2.0}$ & $30.1^{+2.9}_{-2.1}$ & $0.19^{+0.06}_{-0.07}$ & $3.10E-04$ & 1.00  \\
    $\rm GW190719\_215514$ & $36.6^{+42.1}_{-11.1}$ & $19.9^{+10.0}_{-9.3}$ & $54.5^{+38.3}_{-11.1}$ & $0.61^{+0.39}_{-0.30}$ & $6.30E-01$ & 0.92  \\
    $\rm GW190720\_000836$ & $14.2^{+5.6}_{-3.3}$ & $7.5^{+2.2}_{-1.8}$ & $20.8^{+3.9}_{-2.0}$ & $0.16^{+0.11}_{-0.05}$ & $1.00E-05$ & 1.00  \\
    $\rm GW190725\_174728$ & $11.8^{+10.1}_{-3.0}$ & $6.3^{+2.1}_{-2.5}$ & $17.6^{+7.7}_{-1.8}$ & $0.20^{+0.09}_{-0.08}$ & $4.60E-01$ & 0.96  \\
    $\rm GW190727\_060333$ & $38.9^{+8.9}_{-6.0}$ & $30.2^{+6.5}_{-8.3}$ & $65.4^{+9.5}_{-7.3}$ & $0.52^{+0.18}_{-0.18}$ & $1.00E-05$ & 1.00  \\
    $\rm GW190728\_064510$ & $12.5^{+6.9}_{-2.3}$ & $8.0^{+1.7}_{-2.6}$ & $19.7^{+4.4}_{-1.4}$ & $0.18^{+0.05}_{-0.07}$ & $1.00E-05$ & 1.00  \\
    $\rm GW190731\_140936$ & $41.8^{+12.7}_{-9.1}$ & $29.0^{+10.2}_{-9.9}$ & $67.4^{+15.3}_{-10.8}$ & $0.56^{+0.31}_{-0.26}$ & $3.30E-01$ & 0.83  \\
    $\rm GW190803\_022701$ & $37.7^{+9.8}_{-6.7}$ & $27.6^{+7.6}_{-8.5}$ & $62.1^{+11.2}_{-7.6}$ & $0.54^{+0.22}_{-0.22}$ & $7.30E-02$ & 0.97  \\
    $\rm GW190805\_211137$ & $46.2^{+15.4}_{-11.2}$ & $30.6^{+11.8}_{-11.3}$ & $72.4^{+18.2}_{-13.2}$ & $0.92^{+0.43}_{-0.40}$ & $6.30E-01$ & 0.95  \\
    $\rm GW190828\_063405$ & $31.9^{+5.4}_{-4.1}$ & $25.8^{+4.9}_{-5.3}$ & $54.3^{+7.3}_{-4.0}$ & $0.38^{+0.10}_{-0.15}$ & $1.00E-05$ & 1.00  \\
    $\rm GW190828\_065509$ & $23.7^{+6.8}_{-6.7}$ & $10.4^{+3.8}_{-2.2}$ & $33.0^{+5.3}_{-4.3}$ & $0.29^{+0.11}_{-0.11}$ & $3.50E-05$ & 1.00  \\
    $\rm GW190910\_112807$ & $43.8^{+7.6}_{-6.8}$ & $34.2^{+6.6}_{-7.3}$ & $74.4^{+8.5}_{-8.6}$ & $0.29^{+0.17}_{-0.11}$ & $2.90E-03$ & 1.00  \\
    $\rm GW190915\_235702$ & $32.6^{+8.8}_{-4.9}$ & $24.5^{+4.9}_{-5.8}$ & $54.7^{+6.6}_{-5.0}$ & $0.32^{+0.11}_{-0.11}$ & $1.00E-05$ & 1.00  \\
    $\rm GW190916\_200658$ & $43.8^{+19.9}_{-12.6}$ & $23.3^{+12.5}_{-10.0}$ & $65.0^{+17.3}_{-12.6}$ & $0.77^{+0.45}_{-0.32}$ & $2.90E-03$ & 1.00  \\
    $\rm GW190924\_021846$ & $8.8^{+4.3}_{-1.8}$ & $5.1^{+1.2}_{-1.5}$ & $13.3^{+3.0}_{-0.9}$ & $0.11^{+0.04}_{-0.04}$ & $1.00E-05$ & 1.00  \\
    $\rm GW190925\_232845$ & $20.8^{+6.5}_{-2.9}$ & $15.5^{+2.5}_{-3.6}$ & $34.9^{+3.5}_{-2.6}$ & $0.19^{+0.08}_{-0.07}$ & $7.20E-03$ & 0.99  \\
    $\rm GW190926\_050336$ & $41.1^{+20.8}_{-12.5}$ & $20.4^{+11.4}_{-8.2}$ & $59.6^{+22.1}_{-11.8}$ & $0.55^{+0.44}_{-0.26}$ & $1.10E+00$ & 0.54  \\
    $\rm GW190929\_012149$ & $66.3^{+21.6}_{-16.6}$ & $26.8^{+14.7}_{-10.6}$ & $90.3^{+22.3}_{-14.6}$ & $0.53^{+0.33}_{-0.20}$ & $1.60E-01$ & 0.87  \\
    $\rm GW190930\_133541$ & $14.2^{+8.0}_{-4.0}$ & $6.9^{+2.4}_{-2.1}$ & $20.2^{+6.1}_{-2.0}$ & $0.16^{+0.06}_{-0.06}$ & $1.00E-05$ & 1.00  \\
    $\rm GW191103\_012549$ & $11.8^{+6.2}_{-2.2}$ & $7.9^{+1.7}_{-2.4}$ & $19.0^{+3.8}_{-1.7}$ & $0.20^{+0.09}_{-0.09}$ & $4.60E-01$ & 0.96  \\
    $\rm GW191105\_143521$ & $10.7^{+3.7}_{-1.6}$ & $7.7^{+1.4}_{-1.9}$ & $17.6^{+2.1}_{-1.2}$ & $0.23^{+0.07}_{-0.09}$ & $1.20E-02$ & 0.99  \\
    $\rm GW191109\_010717$ & $65.0^{+11.0}_{-11.0}$ & $47.0^{+15.0}_{-13.0}$ & $107.0^{+18.0}_{-15.0}$ & $0.25^{+0.18}_{-0.12}$ & $1.80E-04$ & 0.99  \\
    $\rm GW191113\_071753$ & $29.0^{+12.0}_{-14.0}$ & $5.9^{+4.4}_{-1.3}$ & $34.0^{+11.0}_{-10.0}$ & $0.26^{+0.18}_{-0.11}$ & $2.60E+01$ & 0.68  \\
    $\rm GW191126\_115259$ & $12.1^{+5.5}_{-2.2}$ & $8.3^{+1.9}_{-2.4}$ & $19.6^{+3.5}_{-2.0}$ & $0.30^{+0.12}_{-0.13}$ & $1.00E-05$ & 1.00  \\
    $\rm GW191127\_050227$ & $53.0^{+47.0}_{-20.0}$ & $24.0^{+17.0}_{-14.0}$ & $76.0^{+39.0}_{-21.0}$ & $0.57^{+0.40}_{-0.29}$ & $2.50E-01$ & 0.74  \\
    $\rm GW191129\_134029$ & $10.7^{+4.1}_{-2.1}$ & $6.7^{+1.5}_{-1.7}$ & $16.8^{+2.5}_{-1.2}$ & $0.16^{+0.05}_{-0.06}$ & $1.20E-02$ & 0.99  \\
    $\rm GW191204\_110529$ & $27.3^{+10.8}_{-5.9}$ & $19.2^{+5.5}_{-6.0}$ & $45.0^{+8.7}_{-7.5}$ & $0.34^{+0.25}_{-0.18}$ & $3.30E+00$ & 0.74  \\
    $\rm GW191204\_171526$ & $11.7^{+3.3}_{-1.7}$ & $8.4^{+1.3}_{-1.7}$ & $19.2^{+1.7}_{-0.9}$ & $0.13^{+0.04}_{-0.05}$ & $1.00E-05$ & 0.99  \\
    $\rm GW191215\_223052$ & $24.9^{+7.1}_{-4.1}$ & $18.1^{+3.8}_{-4.1}$ & $41.4^{+5.1}_{-4.1}$ & $0.35^{+0.13}_{-0.14}$ & $1.00E-05$ & 0.99  \\
    $\rm GW191216\_213338$ & $12.1^{+4.6}_{-2.2}$ & $7.7^{+1.6}_{-1.9}$ & $18.9^{+2.8}_{-0.9}$ & $0.07^{+0.02}_{-0.03}$ & $1.00E-05$ & 1.00  \\
    $\rm GW191222\_033537$ & $45.1^{+10.9}_{-8.0}$ & $34.7^{+9.3}_{-10.5}$ & $75.5^{+15.3}_{-9.9}$ & $0.51^{+0.23}_{-0.26}$ & $1.00E-05$ & 0.99  \\
    $\rm GW191230\_180458$ & $49.4^{+14.0}_{-9.6}$ & $37.0^{+11.0}_{-12.0}$ & $82.0^{+17.0}_{-11.0}$ & $0.69^{+0.26}_{-0.27}$ & $5.00E-02$ & 0.96  \\
    $\rm GW200112\_155838$ & $35.6^{+6.7}_{-4.5}$ & $28.3^{+4.4}_{-5.9}$ & $60.8^{+5.3}_{-4.3}$ & $0.24^{+0.07}_{-0.08}$ & $1.00E-07$ & 1.00  \\
    $\rm GW200128\_022011$ & $42.2^{+11.6}_{-8.1}$ & $32.6^{+9.5}_{-9.2}$ & $71.0^{+16.0}_{-11.0}$ & $0.56^{+0.28}_{-0.28}$ & $4.30E-03$ & 0.99  \\
    $\rm GW200129\_065458$ & $34.5^{+9.9}_{-3.1}$ & $29.0^{+3.3}_{-9.3}$ & $60.2^{+4.1}_{-3.2}$ & $0.18^{+0.05}_{-0.07}$ & $1.00E-05$ & 0.99  \\
    $\rm GW200202\_154313$ & $10.1^{+3.5}_{-1.4}$ & $7.3^{+1.1}_{-1.7}$ & $16.8^{+1.9}_{-0.7}$ & $0.09^{+0.03}_{-0.03}$ & $1.00E-05$ & 0.99  \\
    $\rm GW200208\_130117$ & $37.7^{+9.3}_{-6.2}$ & $27.4^{+6.3}_{-7.3}$ & $62.5^{+7.5}_{-6.4}$ & $0.40^{+0.15}_{-0.14}$ & $7.30E-02$ & 0.97  \\
    $\rm GW200208\_222617$ & $51.0^{+103.0}_{-30.0}$ & $12.3^{+9.2}_{-5.5}$ & $61.0^{+99.0}_{-26.0}$ & $0.66^{+0.53}_{-0.29}$ & $4.80E+00$ & 0.70  \\
    $\rm GW200209\_085452$ & $35.6^{+10.5}_{-6.8}$ & $27.1^{+7.8}_{-7.8}$ & $59.9^{+13.1}_{-8.9}$ & $0.57^{+0.25}_{-0.26}$ & $1.00E-07$ & 1.00  \\
    $\rm GW200216\_220804$ & $51.0^{+22.0}_{-13.0}$ & $30.0^{+14.0}_{-16.0}$ & $78.0^{+19.0}_{-13.0}$ & $0.63^{+0.37}_{-0.29}$ & $4.80E+00$ & 0.70  \\
    $\rm GW200219\_094415$ & $37.5^{+10.1}_{-6.9}$ & $27.9^{+7.4}_{-8.4}$ & $62.2^{+11.7}_{-7.8}$ & $0.57^{+0.22}_{-0.22}$ & $9.90E-04$ & 0.99  \\
    $\rm GW200220\_061928$ & $87.0^{+40.0}_{-23.0}$ & $61.0^{+26.0}_{-25.0}$ & $141.0^{+51.0}_{-31.0}$ & $0.90^{+0.55}_{-0.40}$ & $6.80E+00$ & 0.62  \\
    $\rm GW200220\_124850$ & $38.9^{+14.1}_{-8.6}$ & $27.9^{+9.2}_{-9.0}$ & $64.0^{+16.0}_{-11.0}$ & $0.66^{+0.36}_{-0.31}$ & $1.00E-05$ & 1.00  \\
    $\rm GW200224\_222234$ & $40.0^{+6.7}_{-4.5}$ & $32.7^{+4.8}_{-7.2}$ & $68.7^{+6.7}_{-4.8}$ & $0.32^{+0.08}_{-0.11}$ & $1.00E-05$ & 0.99  \\
    $\rm GW200225\_060421$ & $19.3^{+5.0}_{-3.0}$ & $14.0^{+2.8}_{-3.5}$ & $32.1^{+3.5}_{-2.8}$ & $0.22^{+0.09}_{-0.10}$ & $1.10E-05$ & 0.99  \\
    $\rm GW200302\_015811$ & $37.8^{+8.7}_{-8.5}$ & $20.0^{+8.1}_{-5.7}$ & $55.5^{+8.9}_{-6.6}$ & $0.28^{+0.16}_{-0.12}$ & $1.10E-01$ & 0.91  \\
    $\rm GW200306\_093714$ & $28.3^{+17.1}_{-7.7}$ & $14.8^{+6.5}_{-6.4}$ & $41.7^{+12.3}_{-6.9}$ & $0.38^{+0.24}_{-0.18}$ & $2.40E+01$ & 0.81  \\
    $\rm GW200308\_173609$ & $60.0^{+166.0}_{-29.0}$ & $24.0^{+36.0}_{-13.0}$ & $88.0^{+169.0}_{-47.0}$ & $1.04^{+1.47}_{-0.57}$ & $2.40E+00$ & 0.86  \\
    $\rm GW200311\_115853$ & $34.2^{+6.4}_{-3.8}$ & $27.7^{+4.1}_{-5.9}$ & $59.0^{+4.8}_{-3.9}$ & $0.23^{+0.05}_{-0.07}$ & $1.00E-05$ & 0.99  \\
    $\rm GW200316\_215756$ & $13.1^{+10.2}_{-2.9}$ & $7.8^{+2.0}_{-2.9}$ & $20.2^{+7.4}_{-1.9}$ & $0.22^{+0.08}_{-0.08}$ & $1.00E-05$ & 0.99  \\
    $\rm GW200322\_091133$ & $38.0^{+130.0}_{-22.0}$ & $11.3^{+24.3}_{-6.0}$ & $48.0^{+132.0}_{-22.0}$ & $0.59^{+1.43}_{-0.32}$ & $1.40E+02$ & 0.62  \\
\end{longtable}
\footnotesize
    {Note. The data are taken from the GWTC catalogue (https://gwosc.org/).
    Column 1 represents the name of the GW events. Columns 2, 3 and 4 represent
    the source masses of the heavier, lighter and remnant object of the mergers.
    The redshift is listed in Column 5. Columns 6 and 7 represent the false
    alarm rate (FAR) and the corresponding probability of being astrophysical origin.}

\end{document}